\documentstyle[12pt,aasms4]{article} 
\slugcomment{To be submitted to the Astronomical Journal, May 1998} 
\lefthead{MOA Collaboration} 
\righthead{Halo of  IC5249} 
\begin{document} 
 
\title{Observation of the halo of the edge-on galaxy IC\,5249}

\author{F. Abe\altaffilmark{1}, I.A. Bond\altaffilmark{2,3}, B.S. 
Carter\altaffilmark{4}, R.J. Dodd\altaffilmark{4}, M. Fujimoto\altaffilmark{5}, J.B. Hearnshaw\altaffilmark{3}, M. Honda\altaffilmark{6}, J. Jugaku\altaffilmark{7}, S. Kabe\altaffilmark{8}, P.M. Kilmartin\altaffilmark{2,3}, B.S. Koribalski\altaffilmark{9}, M. Kobayashi\altaffilmark{8}, K. Masuda\altaffilmark{1}, Y. Matsubara\altaffilmark{1}, M. Miyamoto\altaffilmark{10}, Y. Muraki\altaffilmark{1}, T. Nakamura\altaffilmark{11}, G.R. Nankivell\altaffilmark{4}, S. Noda\altaffilmark{1}, G.S. Pennycook\altaffilmark{2}, L.Z. Pipe\altaffilmark{2}, N.J. 
Rattenbury\altaffilmark{2}, M. Reid\altaffilmark{12}, N.J. Rumsey\altaffilmark{4}, H. Sato\altaffilmark{11}, S. Sato\altaffilmark{5}, To. Saito\altaffilmark{13}, M. Sekiguchi\altaffilmark{6}, D.J. Sullivan\altaffilmark{12}, T. Sumi\altaffilmark{1}, Y. Watase\altaffilmark{8}, T. Yanagisawa\altaffilmark{1}, P.C.M. Yock\altaffilmark{2}, M. Yoshizawa\altaffilmark{10}}

\affil{\textbf{                 (The MOA Collaboration)}}

\altaffiltext{1}{Solar-Terrestrial Environment Laboratory, Nagoya University, 
Nagoya 464, 
Japan}
\altaffiltext{2}{Faculty of Science, University of Auckland, Auckland, New 
Zealand}
\altaffiltext{3}{Department of Physics and Astronomy, University of Canterbury, 
Christchurch, New Zealand}
\altaffiltext{4}{Carter National Observatory, Wellington, New Zealand}
\altaffiltext{5}{Department of Astrophysics, Nagoya University, Nagoya 464, Japan}
\altaffiltext{6}{Institute for Cosmic Ray Research, University of Tokyo, Tokyo 
188, Japan}
\altaffiltext{7}{Research Institute of Civilization, Tokai University, Hiratsuka 
259-12, 
Japan}
\altaffiltext{8}{High Energy Accelerator Research Organization (KEK), Tsukuba 
305, Japan}
\altaffiltext{9}{Australia Telescope National Facility, CSIRO, P.O. Box 76, Epping, NSW 2121, Australia}
\altaffiltext{10}{National Astronomical Observatory, Mitaka, Tokyo 181, Japan}
\altaffiltext{11}{Department of Physics, Kyoto University, Kyoto 606, Japan}
\altaffiltext{12}{Department of Physics, Victoria University, Wellington, New 
Zealand}
\altaffiltext{13}{Tokyo Metropolitan College of Aeronautics, Tokyo, Japan}

\clearpage

\begin{abstract} 
We report optical photometry and H\textsc{i} synthesis observations of the southern edge-on $\rm S_{\rm c,d}$ galaxy IC\,5249. The observations were carried out primarily to determine if IC\,5249 has a red halo surrounding it similar to the halo that has recently been found surrounding the northern edge-on galaxy NGC~5907. The galaxy IC\,5249 is almost perfectly edge-on, and it shows no direct sign of warping or flaring out to its optical radius. It has a low luminosity and a slow rotation velocity. We find that it has a remarkably similar halo to NGC~5907. The halo is visible at a distance of 3 kpc from the disk of the galaxy, and at 6 kpc the surface brightness is $\approx 28$ mag arcsec$^{-2}$. The measured distribution of the halo light is similar to the distribution expected for dark matter.

Red dwarfs and red giants are possible sources of the halo light. If red dwarfs are the source, and if they are distributed like dark matter, then their density at galactocentric radius $r$ = 8.5 kpc is $(7 \pm 3) \times$ the local Galactic density of halo red dwarfs. Their total contribution to the mass of IC\,5249 would be about 10\% of the mass of the galaxy. If red giants are the dominant source of the observed halo light, and if they are distributed like Galactic globular clusters, then their density at $r$ = 8.5 kpc is $1.0^{+1.0}_{-0.5}\,\times$ the local Galactic density of halo giants, and they provide a gravitationally insignificant fraction of the mass of the galaxy. 

Further optical measurements are required to identify the source of the halo light in IC\,5249, and also to determine more precisely its distribution. Further radio measurements are required to examine the H\textsc{i} distribution and rotation velocity beyond the optical radius of the galaxy. 
\end{abstract} 
 
 
\clearpage

\section{Introduction} 

Observations of the northern edge-on ${\rm S}_{\rm c}$ galaxy NGC~5907 indicate that the disk of this galaxy is surrounded by a faintly luminous halo (\cite{bar94,mor94,sac94,leq96}). The measured distribution of the halo light is very extended, consistent with an emission distribution $\propto r^{-2}$ outside the core of the galaxy. This is, of course, what is expected for dark matter if it is composed of very faint objects, or traced by a visible component\footnote{This could arise, for example, if halo dark matter is composed predominantly of brown dwarfs and includes an admixture of red dwarfs.}. Thus, the observations of NGC~5907 raise the interesting possibility that halo 'dark matter' may be detectable optically. More recently, further observations of NGC~5907 were reported (\cite{rud97,leq98}). These again indicate an extended, faint halo surrounding the disk of the galaxy, but they highlight the difficulties involved in quantifying its brightness and colour. The possibility that the halo may be a result of a galactic merger that occurred in the past was also raised (\cite{leq98}).  

We have carried out optical and radio observations of the southern edge-on galaxy IC\,5249 to determine the generality of the above observations. IC\,5249 is similar to NGC~5907 in that it is nearly edge-on, and possesses a small central bulge. However, IC\,5249 differs in other aspects. It has a lower surface brightness, and its rotation velocity is lower. We find further that it possesses a considerably thicker disk. Also, it is known that NGC~5907 possesses a pronounced warp (\cite{san76}), which may contribute to light observed far from the disk. It is not presently known if IC\,5249 is warped.  

We note that the tracer technique has recently been used as a probe of intergalactic mass in a manner that is not dissimilar to that followed here for IC\,5249 (\cite{fer98}). We also note that the gravitational microlensing surveys of the Magellanic Clouds that have been carried out by the MACHO and EROS groups provide further motivation for the present observations (\cite{alc97a,pal97}). Flaring and warping of the Galactic disk, and also the presence of a Galactic thick disk, have been postulated as possible causes of the long-duration events detected in these surveys (\cite{gat97,eva97}).  Studies of external edge-on galaxies such as NGC~5907 and IC\,5249 should assist to determine the general prevalence of such features in spiral galaxies. 

\section{Sources of faint halo light}

We estimate here the surface brightness $\mu$ generated by various populations of stars that may be present in the halos of galaxies such as NGC~5907 and 
IC\,5249. We integrate along a line of sight at a perpendicular distance (i.e., "impact parameter") $b$ from the center of the galaxy. The line of sight is chosen so that it does not pass through the disk of the galaxy. Assuming a density of halo matter $\rho \propto r^{-2}$, where $r$ denotes galactocentric radius, and a halo luminosity $L \propto \rho$, then it may be easily shown that ${\mu} \propto b^{-1}$. We choose a convenient value of $b$ at which to normalise the distribution, viz. $b$ = 8.5 kpc. This lies well outside the central bulges of galaxies like IC\,5249 and NGC~5907 (\cite{har88}), and it enables convenient comparison with our galaxy. At $b$ = 8.5 kpc the surface brightness $\mu$ in mag ${\rm arcsec}^{-2}$ can be shown to satisfy 
\begin{equation}
\mu  =  M +32.9 - 2.5 \times \log N.
\end{equation}
Here $N$ is number of halo stars per cubic kpc at $r$ = 8.5 kpc, and $M$ their absolute magnitude. Eq.~1 applies for a spatial distribution of halo stars $\propto r^{-2}$ only. Other distributions are considered below.

We use Eq.~1 to estimate first the expected surface brightness due to halo giants. The density of halo giants in our galaxy at $r$ = 8.5 kpc is known. It is the local density, and is 36 $\pm 7$ per cubic kpc (\cite{mor93}). Assuming a similar density of giants at $r$ = 8.5 kpc in an external galaxy, and $\rho \propto r^{-2}$, the halo surface brightness at $b$ = 8.5 kpc due to them is ${\mu}_{\rm R} \approx 28$ mag ${\rm arcsec}^{-2}$. This follows from Eq.~1, assuming ${\rm M}_{\rm R} \approx -1$ for halo giants. The reported value for NGC~5907 is $\sim 27$ mag ${\rm arcsec}^{-2}$ (\cite{sac94}). It is thus plausible that red giants provide a significant fraction of the halo light of NGC~5907. The mass-to-light ratio of giants, however, excludes them from being a significant source of halo mass. They may trace the dark halo only (\cite{fuc95}).     

The Fourth Catalogue of Nearby Stars was used recently to determine the local luminosity function of halo red dwarfs (\cite{fuc97}). The luminosity function peaks at $M_V \approx 11$. At this value, the density is $\approx 1.6 \times 10^5 $ per cubic kpc. Assuming a similar density at $r$ = 8.5 kpc in an external galaxy, and $\rho \propto r^{-2}$, the predicted surface brightness at $b$ = 8.5 kpc is ${\mu}_{\rm R} \approx 30$ mag $ {\rm arcsec}^{-2}$. Here a colour index V - R $\approx 1$ has been assumed\footnote{Magnitudes are in the Johnson passband system unless stated otherwise.}. The predicted surface brightness is approximately 3 magnitudes dimmer than that reported for NGC~5907. This implies the density of halo red dwarfs in NGC~5907 must be about 10$\times$ greater if they provide the halo light of the galaxy. Such a density would account for a few \% of the mass of the halo (\cite{gra96}).  

The long-duration gravitational microlensing events observed towards the 
Magellanic Clouds may be interpreted in terms of a dense population of cool white dwarfs in the Galactic halo (\cite{gra97}). The implied local density and luminosity of the white dwarfs are $N \approx 10^7$ per cubic kpc and ${\rm M}_{\rm B} > 21$, respectively (\cite{fuc97}). Assuming a colour index B - R $\approx 1$ for this population, Eq.~1 implies ${\mu}_{\rm R} > 35$ mag ${\rm arcsec}^{-2}$ at $b$ = 8.5 kpc. This is much fainter than the surface brightness of NGC~5907. 

Surface brightnesses can also be calculated for other distributions. Morrison assumed a spatial density for halo giants similar to that of Galactic globular clusters, i.e. $\propto r^{-3.5}$ (\cite{mor93}). This yields a surface brightness $\mu \propto b^{-2.5}$ (\cite{rud97}), and a normalisation value that may be shown to be one magnitude dimmer than that given by Eq.~1 at $b$ = 8.5 kpc. The $r^{-3.5}$ distribution appears too peaked to fit the data for NGC~5907 (\cite{sac94}). However, it may be applicable to other galaxies. Direct star counts of halo red dwarfs in the Hubble Deep Field indicate that the Galactic distribution of halo red dwarfs falls as fast as or faster than $r^{-3.5}$, and that the distribution is significantly flattened (\cite{fuc97}). Similarly, surface brightness measurements of M\,31 exhibit a much more rapid decline with $b$ than that measured for NGC~5907 (\cite{pri94}). The relatively extended surface brightness distribution of the halo of NGC~5907 may be a characteristic of galaxies with small bulges. 

The surface brightness of globular clusters in distant galaxies in which they are unresolved may be included. Assuming a distribution of globular clusters similar to that in the Galaxy, it may be shown that their effective surface brightness would be $\mu_R \sim$ 31 mag arcsec$^{-2}$.  

In general, it appears that surface brightness measurements are potentially useful for identifying the compositions and other characteristics of the halos of external galaxies. The intensity of the surface brightness, its distribution and its colour, are all useful characteristics for this purpose.

\section{Previous observations of IC\,5249}

The galaxy IC\,5249 was included in the Southern Sky Redshift Survey 
(\cite{cos91}), and in the Southern Sky Survey (\cite{mat92}). Relevant parameters taken from the surveys are:- type = ${\rm S}_{\rm c}$ or ${\rm S}_{\rm d}$; inclination $\approx 89^{\circ}$; ${\rm m}_{\rm I} \approx 13.1$; central surface brightness ${\mu}_{\rm I} \approx 20.1$ mag ${\rm arcsec}^{-2}$; H\textsc{i} flux $\approx$ 27.85 Jy km s$^{-1}$; recession velocity (optical) $\approx$ 2332 km ${\rm s}^{-1}$; recession velocity (H\textsc{i}) $\approx$ 2364 km ${\rm s}^{-1}$; recession velocity (Tully Fisher) $\approx$ 2320 $\pm$ 100 km ${\rm s}^{-1}$; rotation velocity (H\textsc{i}) $\approx$ 102 km s$^{-1}$.

\section{New H\,I measurements of IC\,5249}

\subsection{ATCA H\,I observations and data reduction}

H\textsc{i} synthesis observations of the galaxy IC\,5249 were obtained with the 6C configuration of the Australia Telescope Compact Array (ATCA)\footnote{The Australia Telescope is funded by the Commonwealth of Australia for operation as a National Facility managed by CSIRO.} on 18 October 1992 by Levasseur, Carignan \& Byun (\cite{lev92}). The total time on source was about 12 hours. The 8 MHz bandwidth centred on 1409 MHz was divided into 512 channels which resulted in a velocity resolution of 3.35 km\,s$^{-1}$.

The data reduction and analysis was carried out with the AIPS software package using mostly standard procedures. The $uv$-data were inspected to find those channels containing H\textsc{i} emission. No 20-cm radio continuum emission was detected at the position of IC\,5249 ($S_{\rm peak} < 1.5$ mJy\,beam$^{-1}$). The brightest radio source we detected in the field is PMN J2248-6456 with a flux density of about 27 mJy (\cite{wri94}). The H\textsc{i} line data were Fourier transformed and then cleaned. Use of 'robust weighting' ({\em robust} = 1) resulted in an rms noise of $\sim$2.5 mJy\,beam$^{-1}$ per channel, and an angular resolution of 8 arcsec. IC\,5249 was the only source detected in H\textsc{i}; neither of the galaxies at a projected distance of only 5 arcmin, AM 2243-650 and IC\,5246, were detected in the given velocity range \footnote{ IC\,5246 is shown in Fig.~4 below. The recession velocities of AM 2243-650 and IC\,5246 are unknown.}.

\subsection{The neutral hydrogen distribution and kinematics}          

Fig.~1 shows the H\textsc{i} distribution (0.\,moment) overlaid onto an optical image of IC\,5249. The H\textsc{i} extent of the galaxy is about 280 arcsec $\times$ 24 arcsec (at $N_{\rm H\textsc{i}} = 9 \times 10^{20}$ atoms~cm$^{-2}$) compared to about 254 arcsec $\times$ 30 arcsec in the optical at $R_{25}$ (see Figs. 6, 8 and 9 below). The major axis position angle is $PA$ = 194$^\circ$. We find an integrated H\textsc{i} flux of $\sim$20 Jy km\,s$^{-1}$, corresponding to an H\textsc{i} mass of about $6 \times 10^9 \rm M_{\odot}$ for IC\,5249 assuming a distance of 36 Mpc. This is considerably lower than the total H\textsc{i} flux (27.85 Jy km\,s$^{-1}$) quoted by Matthewson et al. (1992) using the 64-m Parkes radio telescope, and indicates that there is more extended hydrogen emission which has been filtered out with the ATCA interferometer. The current H\textsc{i} distribution shows no evidence of flaring or warping, but more data are needed to study the faint extended structure beyond the optical $R_{25}$ extent of IC\,5249, where warps usually start (see e.g. Briggs 1990). IC\,5249 may well look like NGC~5907 in H\textsc{i} (Sancisi 1976) once we have achieved similar sensitivity.      

The rotation curve of IC\,5249, which was obtained by averaging the H\textsc{i} velocity field (1.\,moment) along a position angle at $PA$ = 194$^\circ$ (width = 22 arcsec), is displayed in Fig.~2. The systemic velocity of the galaxy is $\approx$ 2365 km\,s$^{-1}$. We observe a linear rise of the rotation velocity within a radius of $r$ = 40 arcsec\footnote{In an edge-on galaxy, each rotation velocity measurement is an average of different contributions along the line of sight, and therefore lower than if the major axis only could be observed. We assume most gas is concentrated in the main body of the galaxy, and that this effect is small.}. Beyond that radius the rotation curve slowly turns over and seems to flatten at about $r$ = 100 arcsec or 17.5 kpc, although more sensitive data are needed to confirm this. The maximum observed rotational velocity is about 100 km\,s$^{-1}$. The total mass of IC\,5249 within the observed H\textsc{i} disk (i.e. $r <$ 140 arcsec or 24.5 kpc) is about $5.7\times 10^{10} \rm M_{\odot}$. This assumes the mass inside 24.5 kpc is distributed approximately spherically.      

\section{Photometric measurements}

IC5249 was observed on the nights of July 4 and 6, 1997 with the 0.6m Boller and 
Chivens telescope at the Mt John University Observatory (MJUO) in New Zealand. 
The telescope had recently been modified, with the installation of a stepping motor drive system and wide-angle optics ($1.3^{\circ}$ field). A CCD camera was used with nine, unthinned $1{\rm k} \times 1{\rm k}$ Texas Instrument chips. 

Full details of the camera and modified telescope were reported previously 
(\cite{abe97}). The CCD chips are arranged in a 3$\times$3 matrix, with spacing between adjacent chips equal to 0.9 $\times$ the width of a chip. The pixel size on the sky is 0.645 arcsec. The chip gain and readout-noise are $\approx 5$ e ${\rm ADU}^{-1}$ and 2 e, respectively\footnote{ADU denotes anolgue to digital unit.}. Two broad-passband filters are used, a 'blue' one with approximately 90 \% transmittance from 400 nm to 630 nm, and a 'red' one with similar transmittance from 630 nm to 1,100 nm. The typical sky background at the observatory is U= 21.9, B=22.6, V=21.5, ${\rm R_c}=20.9$ and ${\rm I_c}=19.1$ mag ${\rm arcsec}^{-2}$, and the typical FWHM seeing is 1.5 to 3 arcsec. The camera and modified telescope are currently used primarily for a search for gravitational microlensing events towards the Magellanic Clouds, and a study of non-point-like microlensing events towards the Galactic bulge (\cite{alc97b}).    
 
The visible diameter of IC\,5249 of 4 arcmin enabled it to be accommodated on a single chip of the camera with a large surrounding area of sky. Only three chips of the camera were used for the present observations, the top-left (N-E) one, the center one, and the bottom-right (S-W) one. All exposures were made using the red filter. Seven 20-minute exposures were made with the galaxy on the N-E chip, seven with the galaxy on the center chip, and seven with the galaxy on the S-W chip\footnote{A similar viewing procedure was employed by Lequeux et al (1998). Exposures on different chips enable the effects of sky non-uniformities to be evaluated, as discussed in Sect. 7.1 below.}. Each of these exposures provided two exposures of sky on the two chips not being used for the galaxy. Another seven 20-minute dark exposures were also made on all the chips. Finally, numerous bias and short exposures were made. The exposures were carried out cyclically to minimise effects of changing air-mass and changing camera temperature during a night. Small offsets of the telescope were made between repeat exposures of the same target. Sky conditions on both the observing nights were photometric, and the seeing was approximately 3 arcsec.     
 
\section{Reduction of photometry} 

\subsection{Frame stacking} 

A frame stacking technique similar to that of Morrison et al. (1994) was used to combine the images of IC\,5249. The procedure has been described in full detail elsewhere (\cite{pen98}), and is only outlined here. 

First, the 20-minute dark frames were compared with the bias frames. It was found that the dark current was negligible.  Consequently, bias frames were subtracted from the galaxy and sky frames. 
 
Flatfields were formed from the fourteen 20-minute sky frames for each chip, using standard routines in IRAF\footnote{IRAF is operated by the Association of Universities for Research in Astronomy for the National Optical Astronomical Observatories (USA) under co-operative agreement with the National Science Foundation.}. First, the brighter regions of stellar images were removed from a frame by applying a threshold cut at 1.3 times its modal value. The frames were then modally scaled, and combined using the sigma clipping procedure of IRAF. This self-consistently removes values that are more than $3\sigma$ from the median. The median of the resultant distribution was used as the flatfield. Examination of the distribution of values about the median for any pixel indicated a flatfield accuracy of about 0.85\%. This is consistent with the expectation based on Poisson statistics and the typical ADU values ($\sim 300$) of the sky frames. 
 
Stellar images on the 20-minute frames generally had FWHMs $\sim 4$ arcsec, somewhat in excess of the typical seeing for the system. This resulted from inadequacies of the tracking system, which is unguided, over the longer-than-normal exposure times employed here. Five of the 20-minute galaxy frames had seeing $>$ 4.4 arcsec. These were removed from the sample, leaving a total of sixteen galaxy frames for further analysis, six with the galaxy on the N-E chip, six on the center chip, and four on the S-W one. These were bias and cosmic-ray subtracted, flatfielded, and stacked into a single frame. Positional alignment was achieved to an accuracy of one pixel using six stars. The rotations, translations and scale changes needed for this were carried out using standard IRAF routines. The effective exposure time for the stacked frame is 5hr 20min. The FWHM seeing is  $\approx$ 4.0 arcsec, and the limiting magnitude is $\sim$ 25.5. All of our photometric measurements of IC\,5249 were derived from this 5h 20min frame. It is shown in Fig.~3. 

Globular clusters in IC\,5249 would be expected to have magnitudes $ M_R \sim$ 24.5. It is not evident from Fig.~3 that there is a clustering of such objects surrounding the galaxy, and we conclude that IC\,5249 may be poor in globular clusters. A similar conclusion was reached previously for the galaxy NGC~5907 (\cite{har88}).    
 
\subsection{Calibration of photometry}    

Following each 20-minute exposure of IC\,5249, a 2-minute exposure was made on the same chip. Five uncrowded stars in these exposures with V magnitudes in the range 12 - 15 and colours 0.8 $<$ (B - V) $<$ 1.8 were used to calibrate the photometry. They are shown in Fig.~4. Photoelectric measurements of these stars in the B,V,R$_c$ and I$_c$ passbands were carried out using the 0.6m Optical Craftsman telescope at MJUO and standard stars in the Cousins' F region. These measurements were found to satisfy the relationship 
\begin{equation}
m_R = m_{instr, red} + 24.80 
\end{equation}
to an accuracy of $\pm$ 0.05 magnitude. Here $m_R$ and $m_{instr, red}$ denote photoelectric and CCD instrumental magnitudes, respectively\footnote{The CCD instrumental magnitude was defined in the normal manner, viz. $-2.5\times \log \, \rm (ADU).$}. We used Eq.~2 to calibrate the photometry of IC\,5249.

\subsection{Scattered light}

Scattered light in a telescope, especially at optical surfaces close to the focal plane, may produce an extended halo surrounding a bright object, and contaminate results such as those presented below. Two checks, similar to those carried out by other groups (\cite{mor94,sac94,leq96}), were made to determine the contribution of scattered light to the halo surrounding IC\,5249. First, the point-spread function for bright stars on the stacked galaxy frame was examined. One is shown in Fig.~5. This falls from its central value by a factor of 1000 (corresponding to 7.5 photometric magnitudes) within a radius of 15 pixels, or 1.7 kpc. This implies that scattered light contributes insignificantly to the observed halo. Second, the galaxy image was convolved with the point-spread function, and the light profiles re-calculated. It was found that they differed insignificantly from the original profiles, as expected if scattered light may be neglected (\cite{pen98}).           

\section{Photometry of IC\,5249} 

\subsection{Photometry of the disk of IC\,5249}

A contour plot of the galaxy derived from the 5h 20min frame is shown in 
Fig.~6. This may be compared with the H\textsc{i} distribution displayed in Fig.~1. No flaring or warping is evident in either image, especially of the magnitude required to account for the long-duration gravitational microlensing events observed towards the Magellanic Clouds (\cite{eva97}). The major and minor axis diameters of the R$_{23.5}$ isophote are 220 and 20 arcsec, respectively. 
The photometry of the disk was fitted by exponential functions in both the radial ($R$) and transverse ($z$) directions, viz., 
\begin{equation} 
\mu (R,z) = \mu (0,0) e^{-R/h_R} e^{-|z| / h_z}.  
\end{equation} 
The exponential in the transverse direction is an approximation to a locally isothermal three dimensional sheet (\cite{kru81}). The parameters $h_R$ and $h_z$ are the radial and disk scale heights, and the previously defined quantity $b$ = $\surd(R^2 + z^2)$. 

In what follows we assume ${\rm H}_{\circ}$ = 65 km ${\rm s}^{-1}{\rm Mpc}^{-1}$. This corresponds to a distance of 36 Mpc to IC\,5249, and a pixel size on the sky of 0.113 kpc at the galaxy.

To evaluate the disk and radial scale heights of IC\,5249, photometry was carried out for three profiles at the positions labelled A,B and C in Fig.~3.  They are at distances of 12.1 kpc, 3.2 kpc and 6.8 kpc, respectively, from the nucleus of the galaxy, and their widths are 3.9 kpc, 6.6 kpc and 7.9 kpc. The widths were chosen to avoid foreground stars where possible. The profiles were divided into rectangular bins 1 pixel wide out to a distance of 2 kpc, and thereafter into bins 5 pixels wide. The surface brightness for each bin was determined from the average ADU value. The zero point was taken from the modal value of sky surrounding the galaxy. 

The above procedure was found to be adequate for ${\mu}_{\rm R} <$ 26 mag ${\rm 
arcsec}^{-2}$, but not at fainter levels. As other groups have found, large-scale non-uniformities are present in the sky at levels $\approx 0.3$ \%, and these affect the photometry at ${\mu}_{\rm R} >$ 26 mag ${\rm arcsec}^{-
2}$ (\cite{leq96}). The non-uniformities may be caused by light from nearby bright stars being reflected by components of the telescope. We found that the raw profiles generally decrease monotonically from 0 to about 7 kpc, and thereafter flatten out, but not necessarily to zero. We corrected for this by setting a local zero point for each wing of each profile at the mean raw value between 7 and 10 kpc.

The validity of the above correction procedure was checked by analysing the data from the three CCD chips separately. These are in quite different positions of the focal plane (see Sect.~5), and the sky non-uniformities are quite different on them. However, the corrected profiles were found to be consistent (\cite{pen98}).  

The profiles for the stacked galaxy frame are shown in Fig.~8. The errors shown there are statistical only. Systematic uncertainties of the calibration procedure described in Sect.~6.2 introduce a further uncertainty (not shown in the figure) of $\pm 0.1$ mag ${\rm arcsec}^{-2}$. The systematic uncertainty dominates from 0 to 2.5 kpc, thereafter statistical uncertainty dominates. Also shown in Fig.~9 is a longitudinal profile constructed in a similar manner. For R$>18$ kpc, the luminosity declines much more rapidly than predicted by the exponential model, especially at the northern end of the galaxy. Comparison with the longitudinal I band profile of Mathewson et al. (1992) indicates that the colour index R - I $\approx$ 0.4 in the disk, very similar to its value for NGC~5907 (\cite{leq96}).

Best fits to the exponential model (Eq.~3) from $z$ = 0.5 - 2.0 kpc were obtained with the following values of the scale heights: $h_z$ = 660 $\pm 20$ pc and $h_ R \sim 12 \pm 1$ kpc\footnote{No single value of $h_R$ can be chosen to fit all the profiles simultaneously. This is evident from Fig.~9 below, which shows that the longitudinal profile is only approximately exponential. However, all profiles were fitted simultaneously with $h_z$ = 660 $\pm 20$ pc. The latter value includes the effect of the seeing (4 arcsec) on the 5h 20m image. The deconvolved (\cite{pen98}) value of $h_z = 620 \pm 40$ pc.}.  These fits are shown in Fig.~8. The scale heights for IC\,5249 are larger than for most disk galaxies, implying IC\,5249 is an optically extended galaxy in both the radial and transverse directions\footnote{The corresponding values for NGC~5907 are 467pc and 5.1 kpc respectively (\cite{mor94}).}. Also, the data exceed the exponential model at distances greater than $\sim 3$ kpc from the disk. This is discussed below.  

\subsection{Photometry of the halo of IC\,5249}

Following the discussion in Sect.~2, best fits to the data from 0.5 - 6 kpc were made assuming two sources of light, a disk component given by Eq.~3, and a halo component $\propto b^{-n}$, as follows 
\begin{equation}
\mu (R,z) = \mu (0,0) e^{-R/h_R} e^{-|z| / h_z} + k \times b^{-n}.  
\end{equation}
Here $b$ = $\surd(R^2 + z^2)$ as before, and $n$ = 1 for a luminous halo distributed like dark matter, and $n$ = 2.5 for a visible halo distributed like Galactic globular clusters. The constant $k$ is determined by the density $N$ and absolute magnitude $M$ of the visible stars in the halo, as discussed in Sect.~2. 

Least squares fits to the data with Eq.~4 are shown in Fig.~8.  The data in the halo are subject to large uncertainties, but it is seen that they are not inconsistent (reduced ${\chi}^2 = 2.5$) with a visible halo distributed like dark matter ($n$=1). The value of $k$ for this model corresponds to a halo luminosity $\mu_R$ = 27.75 mag arcsec$^{-2}$ at $b$ = 8.5 kpc, and to 
\begin{equation}
 M_R -2.5 \log N = - 5.15 \pm 0.6 
\end{equation}
at $r$ = 8.5 kpc. 

If we assume the source stars are red dwarfs with $M_R \approx 10$, equation (5) implies a density $N$ of halo red dwarfs at $r$ = 8.5 kpc of $(1.15 \pm 0.7) \times 10^6$ kpc$^{-3}$. This is (7$\pm$3) $\times$ the local value (\cite{fuc97}). Integrating the $\rho \propto r^{-2}$ distribution out to the currently observed {H\textsc{i} radius of $r$ = 140 arcsec or 24.5 kpc yields a total halo mass in red dwarfs $\sim$ 5.5 $10^9 \rm M_{\odot}$ within this radius. This assumes a typical mass $\sim 0.2 \rm M_{\odot}$ for the red dwarfs (\cite{fuc97}), and is very similar to the mass in H\textsc{i} within $r$= 24.5 kpc (see Sect.~4.2). Each component would account for approximately 10\% of the total mass of the galaxy within this radius, quite similar to the situation in NGC~5907 (see Sects. 2 and 4.2). 

The Galactic globular cluster distribution ($n$ = 2.5) shown in Fig.~8 is also not ruled out by the data, although the fit is marginally worse (reduced ${\chi}^2 = 2.8$). The fitted value of $k$ for this model corresponds to a halo luminosity of 28.95 mag ${\rm arcsec}^{-2}$ at $b$ = 8.5 kpc, and to   
\begin{equation}
M_R  -2.5 \log N = - 4.95 \pm 1.0.
\end{equation} 
at $r$ = 8.5 kpc. 

If we interpret Eq.~6 in terms of red giants with $M_R \approx$ -1, we obtain a density of ${38}^{+40}_{-20} {\rm kpc}^{-3}$ halo giants in IC5249 at r = 8.5 kpc. This is ${1.0}^{+1.0}_{-0.5} \times$ the local Galactic value (\cite{mor93}).

Clearly, the above two interpretations of the present data on IC\,5249 in terms of red dwarfs and red giants do not exhaust all the possibilities. For example, the present data may be able to be interpreted in terms of a main disk and an additional 'very thick' disk (\cite{leq98}). Both of the above interpretations are, however, viable, in the sense that the former is similar to a possible interpretation of the data on the halo of the morphologically similar galaxy NGC~5907 (\cite{sac94}), and the latter is similar to the conventional viewpoint of the Galactic halo. Further discussion of the results is given in the final section.

\section{Discussion and conclusions} 

The previous measurements of the ${\rm S}_{\rm c,d}$ galaxy IC\,5249 that were reported in the Southern Sky Survey and in the Southern Sky Redshift Survey have been extended. The {H\textsc{i} distribution shows no sign of warping out to the currently observed limit of $r$ =140 arcsec or 24.5 kpc. The rotation curve rises linearly out to $r$ = 40 arcsec, and then appears to flatten out at about 100 km s$^{-1}$. The colour, central surface brightness, scale height and scale length of the disk of IC\,5249 are R - I $\approx 0.4$, $\mu_{\rm R}$ = 20.5 mag arcsec$^{-2}$, 620 $\pm$ 40 pc and $\sim 12$ kpc, respectively. Additional light to that predicted by an exponential disk is present at distances greater than 3 kpc from the disk. At 6 kpc the surface brightness is ${\mu}_{R} \approx 28$ mag arcsec$^{-2}$. 

The measured distribution of halo light is consistent with that expected for dark matter with $\rho$ and $L \sim r^{-2}$, but other less extended distributions cannot be ruled out. The distributions of halo light surrounding IC\,5249 and NGC~5907 are remarkably similar, suggesting a common origin. This argues against the suggestion of Lequeux et al. (1998) that the halo light of NGC~5907 may be due to a previous galactic merger, unless such mergers are common. Lequeux et al. (1998) note that the transverse light profiles of NGC~5907 are asymmetric, especially in the B passband. The present R band profiles shown in Fig.~8 for IC\,5249 also exhibit possible asymmetry. Such asymmetries may be due to a variety of causes, including unresolved foreground stars, galactic warp or scattered light in a telescope. Further, higher quality data are required to interpret these asymmetries.       

Red dwarfs and red giants are possible sources of the halo light of IC\,5249. If red dwarfs are the source, and if they are distributed like dark matter with $\rho \sim r^{-2}$, then their spatial density at galactocentric radius $r = 8.5$ kpc is $(1.15 \pm 0.7) \times 10^6$ kpc$^{-3}$, or $(7 \pm 3) \times$ the local density of halo dwarfs. Their total contribution to the mass of IC\,5249 within a radius $r$ = 24.5 kpc would be approximately 10\% of the mass of the galaxy within this radius. A similar conclusion holds for NGC~5907 if its halo light is produced by red dwarfs. These conclusions assume the mass and luminosity functions of halo red dwarfs in IC\,5249 and NGC~5907 are similar to the Galactic mass and luminosity functions.    

Brown dwarfs are likely to comprise a significant fraction of halos if red dwarfs do, depending on the form of the initial stellar mass function at low masses. Graff and Freese (1996) considered a model with a mass function $\sim 1/M$. With this model the total mass $M_{tot}$ in a particular mass range $\Delta$ is proportional to the mass range, $M_{tot} \propto \Delta$. If this applies to IC\,5249 and NGC~5907, then it is possible that significant mass resides in brown dwarfs in both these galaxies.   

If red giants are the dominant source of the observed halo light in IC\,5249, and if they are distributed like Galactic globular clusters with  $\rho \sim r^{-3.5}$, then their density at $r$ = 8.5 kpc is $38^{+40}_{-20}$ kpc$^{-3}$, or $1.0^{+1.0}_{-0.5}\, \times$ the local Galactic density of halo giants, and their contribution to the mass of the halo would be gravitationally negligible.
A qualitatively similar result holds for NGC~5907 if its halo light is produced by red giants.

The present measurements of IC\,5249 are subject to significant uncertainties which allow several interpretations of the data, including some not discussed above. Further measurements would be useful. A more accurate determination of the distribution of halo light may be able to distinguish unambiguously between distributions $\sim r^{-2}$ and those that fall faster, and multi-wavelength photometry should differentiate between red dwarfs and red giants as the major source of halo light. More extended H\textsc{i} measurements to greater galactic radii would determine more completely the distribution of mass in the galaxy, and could also exhibit any warping or flaring that may be contaminating the optical photometry of the halo at large distances from the disk.

In general, we conclude that integrated light measurements of the halos of edge-on spiral galaxies are useful for identifying visible stars in the halos of these galaxies, especially those with less prominent nuclei, and that present measurements are consistent with red dwarfs, and associated brown dwarfs, providing a significant fraction of the mass of these galaxies.
 
\clearpage 

Acknowledgements 
 
The authors thank P Sackett and J Lequeux for discussions, the Department of 
Physics and Astronomy of the University of Canterbury for the provision of telescope time, the Marsden and Science Lottery Funds of NZ, the Ministry of Education, Science, Sports and Culture of Japan, the KEK Laboratory of Japan, and the Research Committee of the University of Auckland for financial support.

\clearpage

\figcaption{The H\textsc{i} distribution of the galaxy IC\,5249 overlaid on an optical image obtained from the Digital Sky Survey (DSS). The contour levels are 0.05, 0.1, 0.2, 0.3, 0.4 and 0.5 Jy\,beam$^{-1}$~km\,s$^{-1}$, and the angular resolution is 8 arcsec. North is to the top and east to the left. (The DSS was produced by the Space Telescope Science Institute and is based on photographic data from the UK Schmidt Telescope, the Royal Observatory Edinburgh, the UK Science and Engineering Research Council, and the Anglo-Australian Observatory.) \label{fig 1}}
\figcaption{H\textsc{i} rotation curve of IC\,5249. On the x axis, negative offsets are roughly towards the south and positive offsets roughly towards the north. \label{fig 2}}
\figcaption{Stacked 5h 20m image of IC\,5249 showing positions of the profiles 
used for subsequent photometry. The orientation is similar to that in Fig.~4 below, but with a $14^{\circ}$ rotation applied to align the galaxy vertically. The grey scale ranges from $\sim$ -4 ADU (white) to $\sim$ +9 ADU (black). \label{fig 3}}
\figcaption{Five stars used to calibrate the photometry of IC\,5249. North is to the top, and east to the right. The V, B, R$_{\rm c}$ and I$_{\rm c}$ magnitudes of stars 1, 2, 3, 4 and 5 are (14.41, 15.22, 13.91, 13.74; 14.03, 14.94, 13.53, 13.09; 12.33, 13.12, 11.90, 11.54; 12.08, 12.94, 11.67, -; 13.29, 15.04, 12.94, -). \label{fig 4}}
\figcaption{Point-spread function of the stacked 5h 20m frame. A base level of 5 
ADU has been added to all values to avoid negative levels. The dashed line lies 
7.5 magnitudes below the peak value. This exceeds the fall in galactic brightness from the central value to values 6 kpc from the disc. \label{fig 
5}} 
\figcaption{Contour plot of IC\,5249. The brightest complete contour is at R$_{\rm c} \approx$ 20.5, and the spacing is 0.5 magnitude. The orientation is the same as in Fig.~3. \label{fig 6}}
\figcaption{Stacked 5h 20m image of IC\,5249 showing masked stars excluded from 
the photometry. The orientation is the same as in Fig~3. \label{fig 7}}
\figcaption{Transverse photometric profiles of IC5249. East is to the right, as in Fig.~3. The solid curve is the best fit 'dark halo model', $\rho \propto r^{-2}$, and the dot-dashed curve is the best fit 'globular cluster model', $\rho \propto r^{-3.5}$. The reduced values of $\chi^2$ for these models are 2.5 and 2.8 respectively. The dashed line represents the exponential disk model (Eq.~3). The anomalous excess in the eastern profile of profile A is visible in Figs. 3 and 7. \label{fig 8}}
\figcaption{Longitudinal profile of IC\,5249 in the R$_{\rm c}$ passband.\label{fig 9}}


\clearpage

\begin{thebibliography}{}
\bibitem[Abe et al. 1997]{abe97} Abe, F., Allen, W., Banks, T., Bond, I., 
Carter, B., Dodd, R., Fujimoto, M., Hayashida, N., Hearnshaw, J., Honda, M., 
Jugaku, J., Kabe, S., Kobayashi, M., Kilmartin, P., Kitamura, A., Love, T., 
Matsubara, Y., Miyamoto, M., Muraki, Y., Nakamura, T., Pennycook, G., Pipe, L., 
Reid, M., Sato, H., Sato, S., Saito, To., Sekiguchi, M., Sullivan, D., Watase, 
Y., Yanagisawa, T., Yock, P., Yoshizawa, M. 1997, Variable Stars and the 
Astrophysical Returns of Microlensing Surveys, Ed. R. Ferlet, J.P., Maillard and 
B. Raban, Paris:Editions Frontieres, 1997, 75
\bibitem[Alcock et al. 1997a]{alc97a} Alcock, C., Allsman, R.A., Alves, D., Axelrod, T.S., Becker, A.C., Bennett, D.P., Cook, K.H., Freeman, K.C., Griest, K., Guern, J., Lehner, M.J., Marshall, S.L., Peterson, B.A., Pratt, M.R., Quinn, P.J., Rodgers, A.W., Stubbs, C.W., Sutherland, W. and Welch, D.L. 1997, \apj, 486, 697
\bibitem[Alcock et al. 1997b]{alc97b} Alcock, C., Allen, W.H., Allsman, R.A., Alves,D., Axelrof, T.S., Banks, T.S., Beaulieu, S.F., Becker, A.C., Becker, R.H., Bennett, D.P., Bond, I.A., Carter, B.S., Cook, K.H., Dodd, R.J., Freeman, K.C., Gregg, M.D., Griest, K., Hearnshaw, J.B., Heller, A., Honda, M., Jugaku, J., Kabe, S., Kaspi, S., Kilmartin, P.M., Kitamura, A., Kovo, O., Lehner, M.J., Love, T.E., Marshall, S.L., Maoz, D., Matsubara, Y., Minniti, D., Miyamoto, M., Muraki, Y., Nakamura, T., Peterson, B.A., Pratt, M.R., Quinn, P.J., Reid, I.N., Reid, M., Reiss, D., Retter, A., Rodgers, A.W., Sargent, W.L.W., Sato, H., 
Sekiguchi, M., Stetson, P.B., Stubbs, C.W., Sullivan, D.J., Sutherland, W., Tomaney, A., Vandehei, T., Watase, Y., Welch, D.L., Yanagisawa, T., Yoshizawa, M. and Yock, P.C.M. 1997, \apj, 491,436
\bibitem[Barnaby and Thronson 1994]{bar94} Barnaby, D. and Thronson, H.A. 1994, 
\aj, 107, 1717
\bibitem[Briggs 1990]{bri90} Briggs, F.H. 1990, \aj, 352,15
\bibitem[Costa et al. 1991]{cos91} Costa, L.N., Pellegrini, P.S., Davis, M., 
Sargent, W.L.W. and Tonry, J.L. 1991, \apjs, 75, 935
\bibitem[Evans et al. 1997]{eva97} Evans, N.W., Gyuk, G., Turner, M.S. and 
Binney, J. 1997, astro-ph 9711224, to be published in Astrophysical Journal Letters   
\bibitem[Ferguson et al. 1998]{fer98} Ferguson, H.C., Tanvir, N.R., von Hippel, 
T. 1998, Nature, 391, 461
\bibitem[Fuchs 1995]{fuc95} Fuchs, B. 1995, A\&A, 303, L13
\bibitem[Fuchs 1997]{fuc97} Fuchs, B. 1997, preprint, astro-ph 9708209
\bibitem[Gates et al. 1997]{gat97} Gates, E.I., Gyuk, G., Holder, G.P. and 
Turner, M. 1997, preprint, astro-ph 9711110
\bibitem[Graff and Freese 1996]{gra96} Graff, D.S. and Freese, K. 1996, \apjl, 
456, L49
\bibitem[Graff and Freese 1997]{gra97} Graff, D.S. and Freese, K. 1997, 
preprint, astro-ph 9704125
\bibitem[Harris et al. 1988]{har88} Harris, H.G., Bothun, G.D. and Hesser, J.E. in IAU Symp. No. 126 (eds. Grindlay, J.E. and Davis Philip, A.G., Reidel Dordrecht, 1988),613-614
\bibitem[Kruit and Searle 1981]{kru81} van der Kruit, P.C. and Searle, L. 1981, 
A\&A, 95, 
105
\bibitem[Levasseur et al. 1992]{lev92} Levasseur, P., Carignan, C. and Byun, Y.-I. 1992, ATNF Project C212
\bibitem[Lequeux et al. 1996]{leq96} Lequeux, J., Fort, B., Dantel-Fort, M., 
Cuillandre, J.C. and Mellier, Y. 1996, A\&A, 296, L13
\bibitem[Lequeux et al. 1998]{leq98} Lequeux, J., Combes, F., Dantel-Fort, M., 
Cuillandre, J.C., Fort, B. and Mellier, Y. 1998, preprint, astro-ph 9804109 
\bibitem[Mathewson et al. 1992]{mat92} Mathewson, D.S., Ford, V.L. and Buchhorn, 
M. 1992, \apjs, 81, 413
\bibitem[Morrison 1993]{mor93} Morrison, H. 1993, \aj, 106, 578
\bibitem[Morrison et al. 1994]{mor94} Morrison, H., Boroson, T.A. and Harding, 
P. 1994, \aj, 108, 1191
\bibitem[Palanque-Delabrouille et al. 1997]{pal97} Palanque-Delabrouille, N., Beaulieu, J.P., Krockenberger, M., Sasselov, D.D., Renault, C., Ferlet, R., Vidal-Madjar, E.M., Prevot, L., Aubourg, E., Bareyre, P., Brehin, S., Coutures, C., de Kat, J., Gros, M., Laurent, B., Lachieze-Rey, M., Lesquoy, E., Magneville, C., Milxztajn, A., Moscoso, L., Queinnec, F., Rich, J., Spiro, M., Vigroux, L., Zylberajch, S., Ansari, R., Cavalier, F., Moniez, M. and Gry, C. 1997, preprint, astro-ph 9710194
\bibitem[Pennycook 1998]{pen98} Pennycook, G.S. 1998, University of Auckland thesis in Nagoya University Report No. 0804466 (ed. Muraki,Y., STE Laboratory, Nagoya University, 1998), 204-310   
\bibitem[Pritchet and van den Bergh 1994]{pri94} Pritchet, C.J. and van den 
Bergh, S. 1994, \aj, 107, 1730
\bibitem[Rudy et al. 1997]{rud97} Rudy, R.J., Woodward, C.E., Hodge, T., 
Fairfield, S.W. and Harker, D.E. 1997, Nature, 387, 159
\bibitem[Sackett et al. 1994]{sac94} Sackett, P.D., Morrison, H.L., Harding, P. 
and Boroson, T.A. 1994, Nature, 370, 441
\bibitem[Sancisi 1976]{san76} Sancisi, R. 1976, A\&A, 53, 159
\bibitem[Wright et al. 1994] {wri94} Wright, A.E., Griffith, M.R., Burke, B.F. and Ekers, R.D. 1994, \apjs, 91, 111
\end{thebibliography}
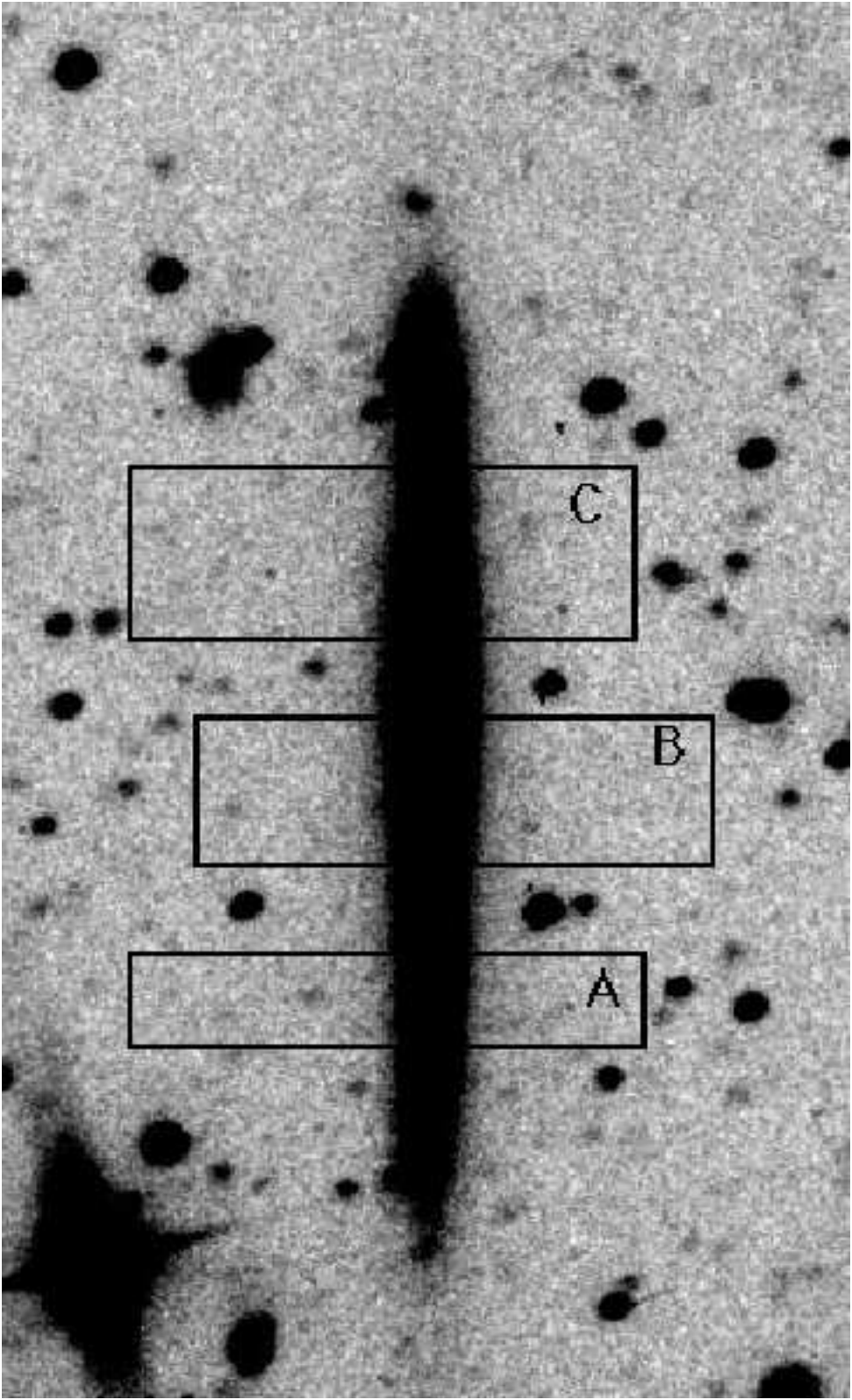
\end{document}